\documentclass[11pt]{article}
\usepackage[english]{babel}
\usepackage[utf8]{inputenc}
\usepackage{latexsym}
\usepackage{amsfonts}
\usepackage{graphicx}
\usepackage{color}
\usepackage{crayola}
\usepackage{xspace}
\usepackage{algpseudocode} 
\usepackage{hyperref}
\usepackage{algorithm}

\newif\ifdraft \newif\ifblind
\draftfalse 
\blindfalse
\ifblind\else\fi

\title{\textbf{Cores in multiway networks}}
\ifblind\author{Author\\ Institution\\ e-mail}\else
\author{Vladimir Batagelj\\IMFM Ljubljana and IAM UP Koper\\ 
e-mail: \texttt{vladimir.batagelj@uni-lj.si}\\ ORCID: 0000-0002-0240-9446}
\fi

\newcommand{\clock}{\count254=\time \divide\count254 by 60
 \count255=\count254 \multiply\count255 by -60
 \advance\count255 by \time
 \ifnum\count254<10 0\fi\number\count254\,:\,%
 \ifnum\count255<10 0\fi\number\count255}

\newcommand{\keyw}[1]{\textcolor{red}{\emph{#1}}}

\newcommand{\RR}{\Bbb{R}}

\newcommand{\network}[1]{\mathcal{#1}}
\newcommand{\vertices}[1]{\mathcal{#1}}

\newcommand{\arcs}[1]{\mathcal{#1}}
\newcommand{\Net}{\network{N}}

\newcommand{\Nodes}{\vertices{V}}
\newcommand{\Links}{\arcs{L}}

\newcommand{\core}{\mathop{\rm core}}

\newcommand{\C}{\mathcal{C}}

\newcommand{\Props}{\mathcal{P}}
\newcommand{\Weights}{\mathcal{W}}
\newcommand{\R}{\mathbb{R}}

\newcommand{\Mw}{\mathop{\raisebox{-1.5pt}{\mbox{$\Box$\kern-.55em\raisebox{2.5pt}{{\tiny $r$}}\kern2.9pt}}}}
\newcommand{\Mv}{\mathop{\raisebox{-1.5pt}{\mbox{$\Box$\kern-.55em\raisebox{2.5pt}{{\tiny $h$}}\kern2.9pt}}}}

\graphicspath{{./}}

\ifdraft\date{\today\  at \clock}\else\date{}\fi

\begin{document}

\maketitle

\begin{abstract}
The notion of a core is generalized to multiway networks. To determine the multiway cores, we adapted already-known algorithms for determining the generalized cores in one-mode and two-mode networks. A new node property, node diversity has been introduced. The newly introduced notions are illustrated with application on the multiway networks of  European airports and airlines and Summer Olympic medals till 2016. For the interactive inspection of the results, their 3D layout in X3D is supported. \medskip

\end{abstract}

\section{Introduction}

One of the approaches to determining important parts of the network is based on the assumption that increased activity in a certain part of the network is reflected in the increased density of the network in that part -- a cohesive group. How can we identify these parts of the network? The densest subgraph is a complete subgraph or clique. Unfortunately, the notion of a clique is algorithmically too demanding for larger networks \cite[problem 3]{moon,NP}. Also, for larger subgraphs, the requirement that inside a clique each node is linked to all others is often unrealistic. For this reason, several milder definitions of the concept of a cohesive group have been proposed, such as $n$-cliques, $n$-clans, $n$-clubs, $k$-plexes, $k$-cores, LS sets, lambda sets, etc. \cite{WaFa}. Among them, only for cores, there exists a very efficient algorithm -- linear with respect to the number of network links \cite{GD99,cores}.

The notion of a core was introduced by Seidman \cite{seidman}. Generalized cores are obtained if, instead of the degree, we take some other property of the nodes as a measure of the importance of the nodes \cite{coresB, cores}. It turns out that the generalization works if the chosen property is monotonic. For two-mode networks, we can choose different properties and different threshold values for each of the two sets of nodes \cite{Ahmed, cores2}. The notion of the core can be extended to temporal networks as well \cite{tcores}. In this article, however, we will extend it to multiway networks.

In the SNA literature, we can find some well-known 3-way networks such as CKM physicians innovation (1957) \cite{physicians}, Kapferer tailor shop (1972) \cite{Kapferer}, Krackhardt office CSS (1987) \cite{Krackhardt}, Lazega law firm (2001) \cite{Lazega}, etc. Recently physicists working on complex networks, for example, Manlio De Domenico (2015) \cite{Manlio}, became interested in multiplex (multi-relational) networks. In 1992 Borgatti and Everett \cite{regbm}, following Baker (1986), extended the blockmodeling to general k-way binary networks. In Genova etal. (2022) \cite{students} a 4-way network about Italian student mobility, based on four ways (provinces, universities, programs, years) was analyzed.

In this paper, we will first set up the context and summarize some basic known facts about cores. Next, we will introduce the notion of the core for multiway networks and propose an algorithm for computing cores. We will conclude with examples of its application on two real-life networks.


\section{Cores in networks}


To set up the context we repeat some basic definitions and results from \cite{cores}.

A network  $\Net = (\Nodes, \Links, \Props, \Weights), \; n = |\Nodes|, m = |\Links|$ is based on four sets: the set of nodes $\Nodes$, the set of links $\Links$, the set of node properties $\Props$, and the set of weights $\Weights$. Each link is linking two nodes -- its end-nodes. The function $\mbox{ext}(e)$ returns the set of end-nodes of the link $e$. A link is either directed -- an arc, or undirected -- an edge. A pair of sets $(\Nodes, \Links)$ forms a graph.

For a subset of nodes $\C \subseteq \Nodes$ we denote with
\[ \Links(\C) = \{ e \in \Links : \mbox{ext}(e) \subseteq \C \} \]
the set of links with both end-nodes inside $\C$.

A \keyw{star} $S(v)$ in a node $v \in \Nodes$ consists of all links having the node $v$ as an end-node
 \[ S(v) = \{ e \in \Links : v \in \mbox{ext}(e)  \}  \]

The notion of a $k$-core was introduced by Seidman in 1983 \cite{seidman}.
\begin{figure}
\centerline{\includegraphics[width=80mm]{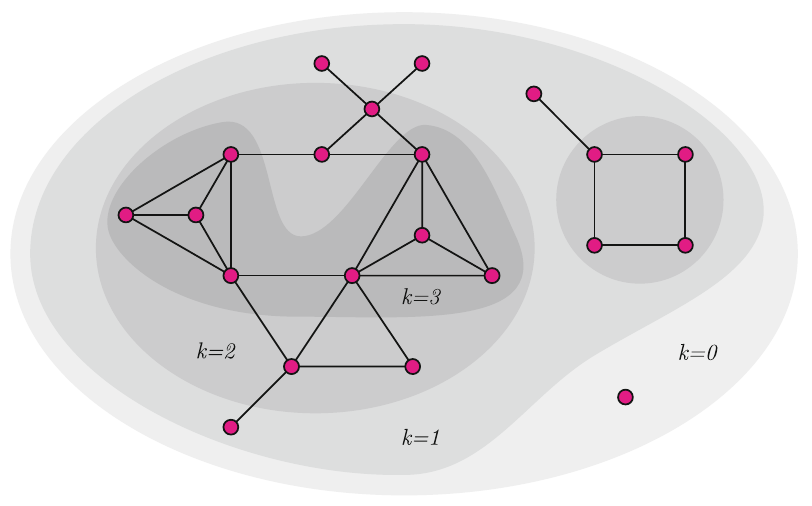} }
\caption{Cores.\label{cores}}
\end{figure}
 Let $\mathcal{G} = (\Nodes, \Links)$ be a graph with set of nodes $\Nodes$ and set of links $\Links$.  The degree of node $v \in \Nodes$ is denoted by $\deg(v)=|S(v)|$. For a given integer $k$, a subgraph $\mathcal{H}_k = (\C_k, \Links(\C_k))$ induced by the subset of nodes $\C_k \subseteq \Nodes$ is called a $k$-core or a core of order $k$  if and only if $\deg_{\mathcal{H}_k}(v) \geq k$, for all $v \in \C_k$, and $\mathcal{H}_k$ is the maximum such subgraph. Usually, also $\C_k$ is called a $k$-core.

The core of maximum order is called the \keyw{main core}.
The \keyw{core number}, $\core(v)$, of node $v$ is the highest order of a core that contains this node.

A  $k$-core graph $\mathcal{H}_k $ is not always connected -- see Figure~\ref{cores}. They are nested
\[ k_1 < k_2 \Rightarrow \mathcal{H}_{k_2}  \subseteq \mathcal{H}_{k_1} \]

Let  $N(v)$ denote the set of neighbors of a node $v$ in a network $\Net$, and $N(v, \C) = N(v) \cap \C$ is the set of neighbors of a node $v$ within the subset of nodes $\C \subseteq \Nodes$. Similarly, 
$S(v,\C) = \{  e \in \Links : v \in \mbox{ext}(e)  \land \mbox{ext}(e) \setminus \{v\} \subseteq \C\} $,
is the set of links incident to node $v$ and having the other node in the set $\C$.

The procedure for determining the $k$-core is simple -- remove from the graph $\mathcal{G}$ all nodes with their current degree less than $k$.\medskip


\begin{algorithmic}[1]
\Function{Core}{$\Net, k$}
\State $\C \gets \Nodes$
\State \textbf{while} $\exists u \in \C : \deg_\C(u) < k$ \textbf{do} $\C \gets \C \setminus \{u\}$
\State \Return $\C$
\EndFunction
\end{algorithmic}
\medskip


A very efficient algorithm exists for determining the \keyw{core decomposition} described with the \keyw{core partition} -- a vector,  $core[v], v \in \Nodes$, of core numbers of all nodes. An  $O(|\Links|)$ implementation of this algorithm is given in \cite{cores}.


\subsection{Generalized cores}

Assume that in a  network $\Net = (\Nodes, \Links, \Props, \Weights)$ a \keyw{node property function} $p(v, \C)$ is defined,  $p: \Nodes \times 2^\Nodes \rightarrow \R$, where $\C \subseteq \Nodes$ is a cluster -- a subset of nodes, and $v \in \C$ is a node.

Some node properties were proposed in \cite{cores}. For example\medskip

\noindent
$p_S (v, \C) = \mathrm{wdeg} (v, \C) =\sum_{u \in S(v, \C)} w(v,u)$ for $w: \Links \rightarrow \mathbb{R}_0^+:$ \keyw{weighted degree}, the sum of weights of incident links within $S(v,\C)$.\medskip

\noindent
They are used in the definition of generalized cores \cite{cores}. 

For a selected node propoerty $p$, the subgraph $\mathcal{H} = (\C,\Links(\C))$ induced by the set $\C \subseteq \Nodes$ is a \keyw{$p$-core at level $t \in \mathbb{R}$} iff $\forall v \in \C : t \leq p(v, \C)$ and $\C$ is the maximum such set. The \keyw{$p$-core value}, $\core(v)$, of  a node $v$ is the highest level of a core that contains this node.

We say that the node property function $p(v, \C)$:
\begin{itemize}
\item  is \keyw{local} iff: $p(v, \C) = p(v, N(v, \C)) \;$ for all $v \in \Nodes$.
\item  is \keyw{monotonic} iff: $\C_1 \subset \C_2 \Rightarrow \forall v \in \Nodes: p(v, \C_1) \leq p(v, \C_2).$
\end{itemize}

For determining the $p$-core at level $t$ we adapt the algorithm for the $k$-core\medskip

\begin{algorithmic}[1]
\Function{CoreG}{$\Net, p,t$}
\State $C \gets \Nodes$
\State \textbf{while} $\exists u \in C : p(u,C) < t$ \textbf{do} $C \gets C \setminus \{u\}$
\State \Return $C$
\EndFunction
\end{algorithmic}\medskip

For a monotonic property function $p$, the result of this algorithm is independent of the nodes' elimination order.

For a local and monotone property function $p$ the corresponding $p$-core values, the generalized core decomposition, can be efficiently determined using a generalized version of the degree core decomposition algorithm \cite{cores}.
If $p(v, N(v, C))$ can be computed in $O(\deg(v, C))$ time then
for a monotone and local node property function $p$ this algorithm determines the $p$-core
hierarchy in $O(m · \max(\Delta, \log n))$ time, $\Delta$ is the maximum degree.



For two-mode networks, the notion of $(k_1, k_2)$-cores was introduced in \cite{Ahmed} in 2007 
and extended to other node property functions in \cite{cores2}.

Let $\Net = ((\Nodes_1, \Nodes_2), \Links, (p_1, p_2), w)$, $\Nodes = \Nodes_1\cup \Nodes_2$ and $\Nodes_1\cap \Nodes_2=\emptyset$ be a two-mode network. Let functions $p_1(v, \C)$ and $p_2(v, \C)$ be node property functions defined on the network  $\Net$.

The subnetwork $Core(p_1, p_2; t_1, t_2) = ((\C_1, \C_2), \Links(\C))$,  $\C_1 = \C \cap \Nodes_1$, $\C_2 = \C \cap \Nodes_2$, and $t_1, t_2 \in \RR^+_0$  induced by the subset of nodes $\C \subseteq \Nodes$ in a two-mode network $\Net$ is a \keyw{generalized two-mode core} for $p_1$ and $p_2$ at levels $t_1$ and $t_2$ if and only if it holds that for all $v \in \C_1 : p_1 (v, \C) \geq t_1$ and for all $v \in \C_2: p_2(v, \C) \geq t_2$, and $\C$ is the maximal such subset in $\Nodes$ \cite{cores2}.

Again, to obtain the generalized two-mode core for node properties $p_1$ and $p_2$ at levels $t_1$ and $t_2$ we can adapt the basic cores algorithm:\medskip

\begin{algorithmic}[1]
\Function{Core2}{$\Net, (p_1, p_2),(t_1, t_2)$}
\State $\C \gets \Nodes$
\State \textbf{while} $\exists u \in \C : (u \in \Nodes_i) \land (p_i(u,\C) < t_i))$ \textbf{do} $\C \gets \C \setminus \{u\}$
\State \Return  $\C$
\EndFunction
\end{algorithmic}\medskip

The two-mode cores are only partially nested
\[  t_1' < t_1 \Rightarrow Core(p_1, p_2; t_1, t_2) \subseteq Core(p_1, p_2; t_1', t_2) \]
\[  t_2' < t_2 \Rightarrow Core(p_1, p_2; t_1, t_2) \subseteq Core(p_1, p_2; t_1, t_2') \]
We can obtain a core decomposition if we fix one of the threshold values. Interesting threshold pairs $(t_1,t_2)$ are the corner points of the corresponding "Pareto stairs" border \cite{cores2}.

\section{Cores and multiway networks}

\subsection{Multiway networks}

A \keyw{weighted multiway network} $\Net = (\Nodes,\Links,w)$ is based on \keyw{nodes} from $k$  finite sets (ways or dimensions) $ \Nodes = ( \Nodes_1, \Nodes_2, \ldots, \Nodes_k ) $, the set of \keyw{links} $\Links \subseteq \Nodes_1\times \Nodes_2\times \cdots\times \Nodes_k$, and the \keyw{weight} $w : \Links \to \RR$. 
If for $i\ne j$, $\Nodes_i = \Nodes_j$, we say that $\Nodes_i$ and $\Nodes_j$ are of the same \keyw{mode}.

In a general multiway network, different additional data can be known for nodes and/or links.

For illustrations of basic notions, we will use a multiway (multi-relational) network $\Net_{MA} = ((\Nodes_{an},\Nodes_{pl},\Nodes_{R}),\Links,w)$ Marmello77 \cite{marmello77} that describes both mutualistic and antagonistic interactions between mammals (animals) and seeds (plants). It has three ways:\medskip
\begin{figure}
{\small
\renewcommand{\baselinestretch}{0.85}
\begin{verbatim}
> MA$links
   an pl R  w       an pl R  w       an pl R  w       an pl R  w
1   5  1 1  2    19  7 13 1  2    37  6 28 1  1    55  7 11 2 12
2   7  2 1 20    20  4 14 1  1    38  7 28 1 18    56  7 12 2  2
3   9  4 1  1    21  4 15 1 50    39  3 29 1  1    57  4 15 2  5
4   2  5 1 13    22  2 16 1  2    40  2 31 1 11    58  2 16 2  1
5   4  5 1  6    23  3 17 1  3    41  7 32 1  2    59  3 17 2  1
6   9  5 1  1    24  3 18 1  5    42  3 33 1  1    60  3 18 2  2
7   2  3 1 12    25  4 19 1  1    43  4 34 1  3    61  4 23 2 36
8   4  3 1 11    26  4 20 1  2    44  7  2 2  1    62  8 23 2  1
9   7  6 1  3    27  9 21 1  1    45  2  5 2  9    63  6 25 2  2
10  1  7 1  6    28  4 22 1  1    46  4  5 2 12    64  7 25 2 62
11  2  7 1 19    29  4 23 1 19    47  2  3 2  3    65  4 27 2  7
12  4  8 1  1    30  2 24 1 72    48  4  3 2  4    66  7 27 2  1
13  4  9 1  3    31  7 24 1 10    49  7  6 2  1    67  2 28 2  2
14  4 10 1  2    32  2 25 1  9    50  1  7 2  1    68  6 28 2  1
15  2 11 1  3    33  7 25 1 58    51  2  7 2  2    69  7 28 2 24
16  7 11 1 15    34  4 26 1  2    52  4  7 2  1    70  3 29 2  3
17  7 12 1  4    35  4 27 1  2    53  4  8 2  6    71  9 30 2  1
18  2 13 1  1    36  7 27 1  9    54  4  9 2  8    72  7 32 2  5
\end{verbatim}
}
\caption{Links with weights in Marmello77 multiway network.\label{Mar77}}
\end{figure}

\noindent
$\Nodes_{an}$ ; \texttt{an} -- \textbf{\textit{Animals}}:     
 (1) CerSco, 
 (2) CerSub, 
 (3) DidAlb, 
 (4) DidAur, 
 (5) GueIng, 
 (6) MonDom, 
 (7) NecLas, 
 (8) OliSp,  
 (9) PhiFre. \medskip

\noindent
$\Nodes_{pl}$ ; \texttt{pl} -- \textbf{\textit{Plants}}:     
 (1) ByrSp,  
 (2) CamAda, 
 (3) CecGla, 
 (4) CecHol, 
 (5) CecPac, 
 (6) ChiAlb, 
 (7) CocAur, 
 (8) ErytSp, 
 (9) EuphSp, 
(10) FicuSp, 
(11) LeaAur, 
(12) LeaCor, 
(13) LeanSp, 
(14) MicAlb, 
(15) MicLig, 
(16) MicPep, 
(17) MicSp1, 
(18) MorSp1, 
(19) MorSp2, 
(20) MorSp3, 
(21) MorSp5, 
(22) MorSp6, 
(23) MorSp7, 
(24) MorSp8, 
(25) MorSp9, 
(26) MyrSp1, 
(27) MyrSp2, 
(28) PsidSp, 
(29) RubSp1, 
(30) RubSp2, 
(31) SabBra, 
(32) SolaSp, 
(33) TocFor, 
(34) VisBra.\medskip

\noindent
$\Nodes_R$ ; \texttt{R} -- \textbf{\textit{Interactions}}:         
(1) antagonistic, 
(2) mutualistic.  \medskip

The list of links $\Links$ and their weight $w$ is presented in Figure~\ref{Mar77}. The weight $w$ is measuring the viability of seeds.

Let $e \in \Links$ be a link. With $e(i)$ we denote the $i$-th component of  $e = (e(1), e(2), \ldots, e(i), \ldots, e(k))$. For example, in the network $\Net_{MA}$ we have $e_{51}(2) = 7 \approx \mbox{CocAur} \in \Nodes_{pl}$. 

For a subset of links $\Links' \subseteq \Links$ we denote 
\[ \Nodes_i(\Links') = \{ e(i) : e \in \Links' \} \]
the set of nodes from the way $\Nodes_i$ that are an end-node of a link from $\Links'$.

For example, for $\Links' = \{ e_i : i \in 51..60 \}$ we have $\Nodes_{an}(\Links') = \{ 2, 3, 4, 7 \} \approx \{$ CerSub, DidAlb, DidAur, NecLas $\}$.

A multiway network $\Net$ is \keyw{simple} if and only if all its links (tuples) are mutually different.

A multiway network $\Net' = ((\Nodes'_1,\Nodes'_2,\ldots,\Nodes'_k),\Links',w')$ is a \keyw{subnetwork} of the multiway network $\Net = ((\Nodes_1,\Nodes_2,\ldots,\Nodes_k),\Links,w)$ if and only if $\Nodes'_i \subseteq \Nodes_i$ for all $i \in 1..k$,  $\Links' \subseteq \Links$, $\Links' \subseteq \Nodes'_1 \times \Nodes'_2 \times \ldots \times \Nodes'_k$, and for all $e \in \Links'$ holds $w'(e) = w(e)$.

The notion of a subnetwork could be extended to the omission of some ways.

\subsection{Node property functions in multiway networks}

For a node $u \in \Nodes_i$ we call a \keyw{star} in the node $u$, the set of links
\[ S(u) = \{ e \in \Links : e(i) = u \}.\]
For example, for $u = 9 \approx \mbox{PhiFre} \in \Nodes_{an}$ we have
$S(\mbox{PhiFre}) = \{ e_3, e_6, e_{27}, e_{71}\}$.

A list of subways $\C = (C_1, C_2, \ldots, C_r)$, $C_i \subseteq \Nodes_{h(i)}$ we will call a \keyw{selection}. 
Note that the selection $\C$ defines a function $h : 1..r \to 1..k$. For example, the selection $\C =  (\Nodes_{pl}, \{ \mbox{mutualistic}\}_R )$ defines the function $h$: $h(1) = 2 \approx \mbox{pl}$, $h(2) = 3 \approx \mbox{R}$.

For a link $e \in \Links$ we introduce abbreviations
\[ h(e) \equiv (e(h(1)), e(h(2)), \ldots, e(h(r)) )\]
for a \keyw{selected part} of the link $e$; and
\[ h(e) \in \C \equiv \forall i \in 1..r : e(h(i)) \in C_i \]
for a statement that all nodes of the selected part of the link $e$ belong to selected subways.

For a node $u \in \Nodes_i$ and a selection $\C$, we call a \keyw{star} in the node $u$ for the selection $\C$, the set
\[S(u,\C) = \{ e \in \Links : e(i) = u \land h(e) \in \C \}.\]
For example, for $\C = ( \{ 6..9 \}_{an}, \{ 2 \}_R) \approx (\{ $ \mbox{MonDom}, \mbox{NecLas}, \mbox{OliSp}, \mbox{PhiFre} $\}_{an}, \{ \mbox{mutualistic} \}_R)  $ and for $u = 28 \approx \mbox{PsidSp} \in \Nodes_{pl}$ we have $S(\mbox{PsidSp}, \C) = \{ e_{68}, e_{69} \}$.

We extend the notation $\Nodes(\Links')$ to selections by
\[ \C(\Links') = (\C_1(\Links'), \C_2(\Links'),\ldots, \C_r(\Links'))\]

For a node $u \in \Nodes_i$ we call the set of \keyw{neigbors} of the node $u$ in the way $\Nodes_j$ for the selection $\C$, the set of nodes $N(u,\C,\Nodes_j) = \Nodes_j(S(u,\C))$. For example, for the set $\C$ from the definition of a star we have $N(\mbox{PsidSp}, \C, \Nodes_{an}) = \{ 6, 7 \} \approx \{  \mbox{MonDom}, \mbox{NecLas}\}$.

Using stars and sets of neighbors we can define some node property functions on multiway networks. For a node $u \in \Nodes_i$ and a selection $\C$

\noindent
\keyw{degree} 
\[ p_d(u,\C) = \mbox{card}(S(u,\C))\]
\keyw{weighted degree} for the weight $w$
\[ p_S(u,\C; w) = \sum_{e \in S(u,\C)} w(e) \]
\keyw{max weight} for the weight $w$
\[ p_M(u,\C; w) = \max_{e \in S(u,\C)} w(e) \]
\keyw{diversity} of the way $\Nodes_j$
\[ p_\delta(u,\C,\Nodes_j) = \mbox{card}(N(u,\C,\Nodes_j))\]
\keyw{contribution} of a (measured) node property/attribute  $a$ on the way $\Nodes_j$
\[ p_c(u,\C,\Nodes_j; a) = \sum_{v \in N(u,\C,\Nodes_j)} a(v) \]

\subsection{Cores in multiway networks}

Let $\C = (C_1, C_2, \ldots, C_k) \subseteq \Nodes$ be a selection, $\mathbf{p}=(p_i)$ be a list of monotonic node property functions over the selected ways, and $\mathbf{t}=(t_i)$ a list of the corresponding thresholds. 

The multiway subnetwork $Core(\mathbf{p},\mathbf{t}) = (\C, \Links(\C))$  induced by the subset of nodes $\C \subseteq \Nodes$ in a multiway network $\Net$ is a \keyw{generalized multiway core} for node property functions $\mathbf{p}$ at levels $\mathbf{t}$  if and only if it holds that for all $v \in \C_i : p_i (v, \C) \geq t_i$  and $\C$ is the maximal such subset in $\Nodes$.

Then for monotonic node property functions $\mathbf{p}$ the generalized core can be determined by the following algorithm \medskip
\newpage
\begin{algorithmic}[1]
\Function{CoreMW}{$\Net, \mathbf{p}, \mathbf{t}$}
\State $\C \gets (\Nodes_{h(1)}, \Nodes_{h(2)},\ldots,\Nodes_{h(r)})$
\State \textbf{while} $\exists u \in C_i : p_i(u,\C) < t_i$ \textbf{do} $C_i \gets C_i \setminus \{u\}$
\State \Return  $\C$
\EndFunction
\end{algorithmic}\medskip

Multiway cores are partially nested
\[  \mathbf{t}' < \mathbf{t} \Rightarrow Core(\mathbf{p},\mathbf{t}) \subseteq Core(\mathbf{p},\mathbf{t}') \]

In an elaboration of this algorithm, we have many options. Again the result does not depend on the node elimination order. Here we present a simple algorithm that is eliminating nodes way-wise.

The core conditions are given in a list $P$. Each entry of the list is a triple $(p, t, args)$ where $p$ is a node property function, $t$ is the corresponding level, and $args$ is a list of ways on which the function $p$ is defined. It is assumed that the node $u$ belongs to the way $args[1]$.
 
{

\begin{algorithm}
\caption{Generalized multiway core for node properties $\mathbf{p}$ at levels $\mathbf{t}$.\label{MWcore}}
\begin{algorithmic}[1]
\Function{MWcore}{$\Net, h, P$}
\State $\C \gets (\Nodes_{h(1)}, \Nodes_{h(2)}, \ldots, \Nodes_{h(r)})$
\Repeat
   \State $exit \gets true$
   \For{$(p,t,args) \in P$}
      \State $R \gets \emptyset$; $i \gets args[1]$
      \For{$u \in C_i$}
         \If{$p(\Net,u,\C) < t$}
            \State $R \gets R \cup \{u\}$; $exit \gets false$
         \EndIf
      \EndFor
      \State $C_i \gets C_i \setminus R$   
   \EndFor
\Until{exit}
\State \Return  $\C$

\EndFunction

\end{algorithmic}
\end{algorithm}
}
For multiway networks with some ten thousand links, our implementation of Algorithm~\ref{MWcore} in R produces the core in some seconds.
 
For large multiway networks, more efficient algorithms can be developed that are recomputing a node property value only when there was a change in the nodes/links on which it depends. For some property functions, additional speed-up can be achieved by updating the old value.

\subsection{R library MWnets}

To support the analysis of multiway networks we are developing in R a package \texttt{MWnets} available at Github/Bavla \cite{MWnets}. Algorithm~\ref{MWcore} is implemented as the function \texttt{MWcore(MN,P)} where \texttt{MN} is a multiway network and \texttt{P} is
a list of core conditions.

The node property function(s) can be selected among
\begin{itemize}
\item $p_d$: \texttt{pDeg(MN,u,cip,C)} 
\item $p_S$: \texttt{pWsum(MN,u,cip,C,weight=))} 
\item $p_M$: \texttt{pWmax(MN,u,cip,C,weight=))} 
\item $p_\delta$: \texttt{pDiv(MN,u,cip,C,way=)} 
\item $p_c$: \texttt{pAttr(MN,u,cip,C,way=,attr=,FUN=sum)} 
\end{itemize}
The user can implement and use in MWcore his/her own node property functions.

\subsection{3D visualization of multiway networks}

A multiway network $\Net$ can be represented by a $k$-dimensional array $W$
\[  W[v_1,v_2,\ldots,v_k] = w(v_1,v_2,\ldots,v_k) \ \mbox{ for } \  (v_1,v_2,\ldots,v_k) \in \Links \]
otherwise $W[v_1,v_2,\ldots,v_k] = 0$.\medskip

To get some insight into the structure of the network and obtained cores we can inspect their 3D layouts in X3D format. They are created using the MWnets function \texttt{mwnX3D}. In producing the layout it is very important to provide "clever" orderings of the ways. For ordinal ways we usually stick to natural ordering; for nominal (categorical) ways the ordering is often determined using clustering or by sorting some important quantity defined on a selected way.

The 3D layouts can be interactively inspected using special viewers such as \textbf{view3dscene} \cite{X3Dview}. 
They can be also included in HTML web pages and inspected (rotate, move, zoom in/out) using a Javascript-based library X3DOM \cite{X3DOM}. Inspecting a 3D layout the user can also dynamically get information about the selected link (shape) by clicking on it. The link's information is provided in the tag \texttt{<Anchor>} attribute \texttt{description} that is enclosing each link shape. The interactive inspection provides the user with much better insight into the network structure than the snapshots included in this paper. The 3D layouts for this paper are available at Github/Bavla \cite{MWnets}. 

The current implementation of the function \texttt{mwnX3D} provides only the basic functionalities -- for a given simple/simplified multiway network displaying three ways by given orderings, coloring each shape, and the volume of the shape representing a link is proportional to its weight. It can be improved to present up to three weights by $x$, $y$, $z$ sizes of a shape. Also, the set of available shapes can be extended from cube and sphere to other shapes. Using a cube as the shape, up to six weights can be encoded as colors of cube faces.



\section{Examples}

\subsection{European airports and flight companies}


The data set was collected by A. Cardillo and his collaborators for the paper \cite{EUair}. The multiway network
\texttt{AirEu2013} contains 450 airports making two ways 	\texttt{airA} (departure airports) and \texttt{airB} (arrival airports),  the third way is \texttt{line} (37 flight companies). There are 7176 links $(from, to, line) \in \Links$ telling that there is a flight 
provided by the company $line$ from the airport $from$ to the airport $to$. Each link has the weight $w = 1$.

\begin{figure}
\includegraphics[width=125mm]{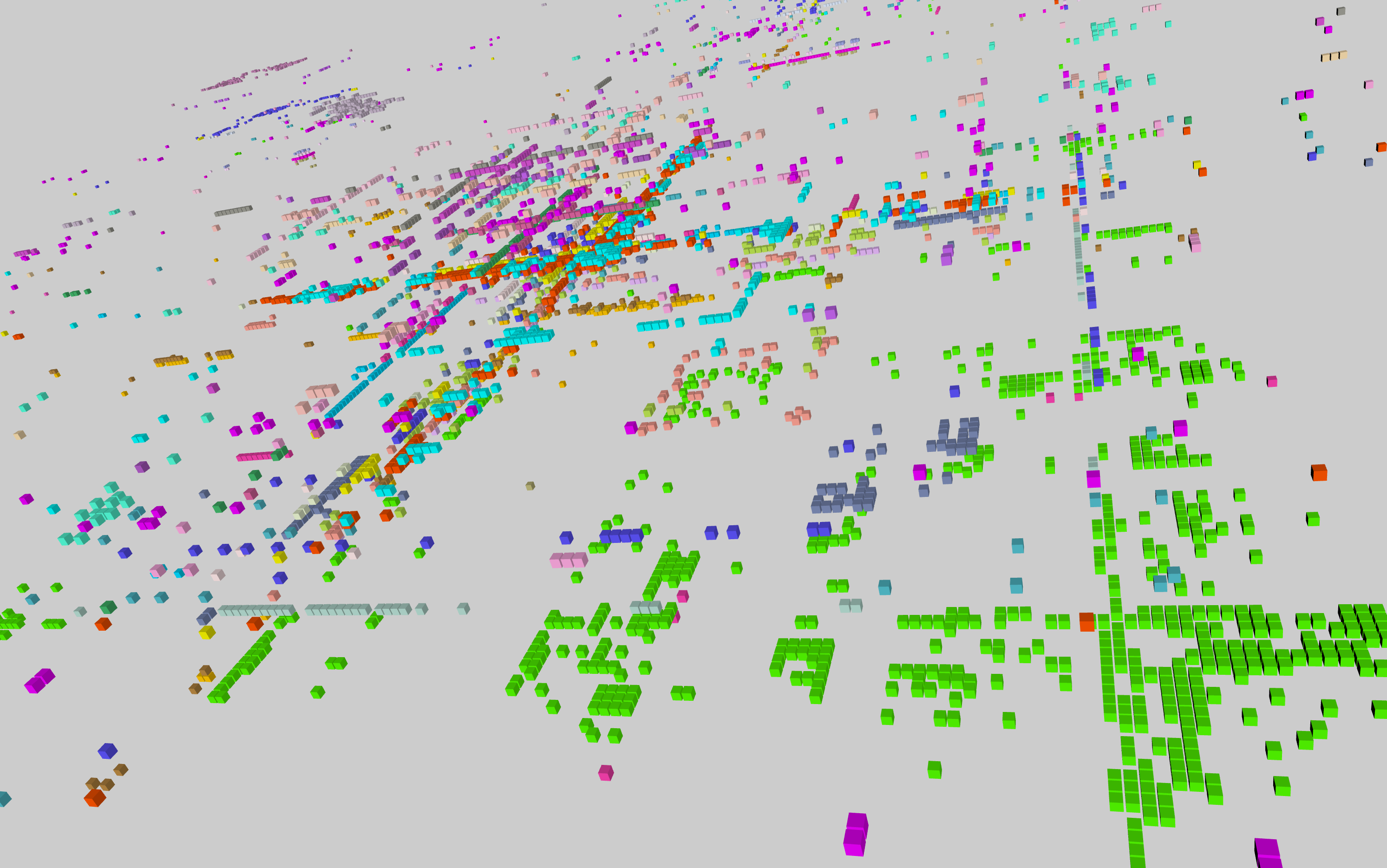}
\caption{Screenshot of a part of the 3D layout of \texttt{AirEu2013}.\label{EUair}}
\end{figure}

In Figure~\ref{EUair} a screenshot of a large part of the 3D layout of \texttt{AirEu2013} is presented.
The ordering of ways airA, airB, and line were determined using clustering.
A cube in the layout represents a link/triple $(ap_1, ap_2, co)$. Lines provided by the same company are of the same color. 
The green links in the bottom right corner are provided by Ryan air. The overlap with other companies is minor -- Ryan air is mostly linking airports that are not linked by other companies.

Let's look for the airports that are linked with a large number of different companies -- on both airport sets we search for cores with a large diversity $pDiv$. The core conditions list $P$ contains two triples $(p=pDiv, t=10, args=(airA, airB, way=line) )$  and
$(p=pDiv, t=10, args=(airB, airA, way=line) )$. The core at level 10 contains 49 airports. Inspecting the obtained core we decided to increase the threshold level to 13. 
We obtained the 13-core on 28 airports -- each airport inside the core is linked to other airports in the core by at least 13 companies.\medskip

\noindent
\textbf{EUair core13 / Airports.} In the list we use abbreviations: 
Ap -- Airport and IAp -- International Airport.\\
 (1) Frankfurt Ap,
 (2) Berlin Tegel Ap,
 (3) Munich Ap,
 (4) Dusseldorf Ap,
 (5) Hamburg Ap,
 (6) Zurich Ap,
 (7) Geneva Ap,
 (8) Milan-Malpensa Ap,
 (9) Copenhagen Ap,
(10) Stockholm Arlanda Ap,
(11) Heathrow Ap,
(12) Warsaw Chopin Ap,
(13) Vienna IAp,
(14) Amsterdam Ap Schiphol,
(15) Charles de Gaulle Ap (Roissy Ap),
(16) Adolfo Suarez Madrid-Barajas Ap,
(17) Barcelona El Prat Ap,
(18) Malaga Ap,
(19) Vaclav Havel Ap Prague,
(20) Leonardo da Vinci-Fiumicino Ap,
(21) Brussels Ap (Zaventem Ap),
(22) Budapest Ferenc Liszt IAp,
(23) Athens IAp (Eleftherios Venizelos Ap),
(24) Ben Gurion Ap,
(25) Henri Coanda IAp,
(26) Venice Marco Polo Ap,
(27) Sofia Ap,
(28) Nice Cote d'Azur Ap.

\noindent
\textbf{EUair core13 / Companies.} Abbreviation A -- Airways.\\
 (1) Lufthansa,  (2)  Air Berlin,   (3) Swiss IAL,   (4) Netjets,   (5) Easyjet,    
 (6) SAS,   (7) Norwegian AS,   (8) British A,   (9) LOT Polish A,   (10) Austrian A,  
(11) Niki,   (12) KLM,   (13) Transavia H,   (14) Air France,   (15) Iberia,      
(16) Air Nostrum,   (17) Vueling A,   (18) Ryanair,   (19) Czech A,   (20) Alitalia,    
(21) Brussels A,   (22) European AT,   (23) Malev HA,   (24)  Wizz Air,   (25) Aegean A    
(26) Olympic Air,   (27) TNT Airways.\medskip

The companies that were not providing any line between core airports are  Turkish A,  Flybe,  TAP Portugal, Finnair,  Air Lingus,  Germanwings,  Pegasus A,  SunExpress,  Air Baltic, and   Wideroe.
  
\begin{figure}
\includegraphics[width=62mm]{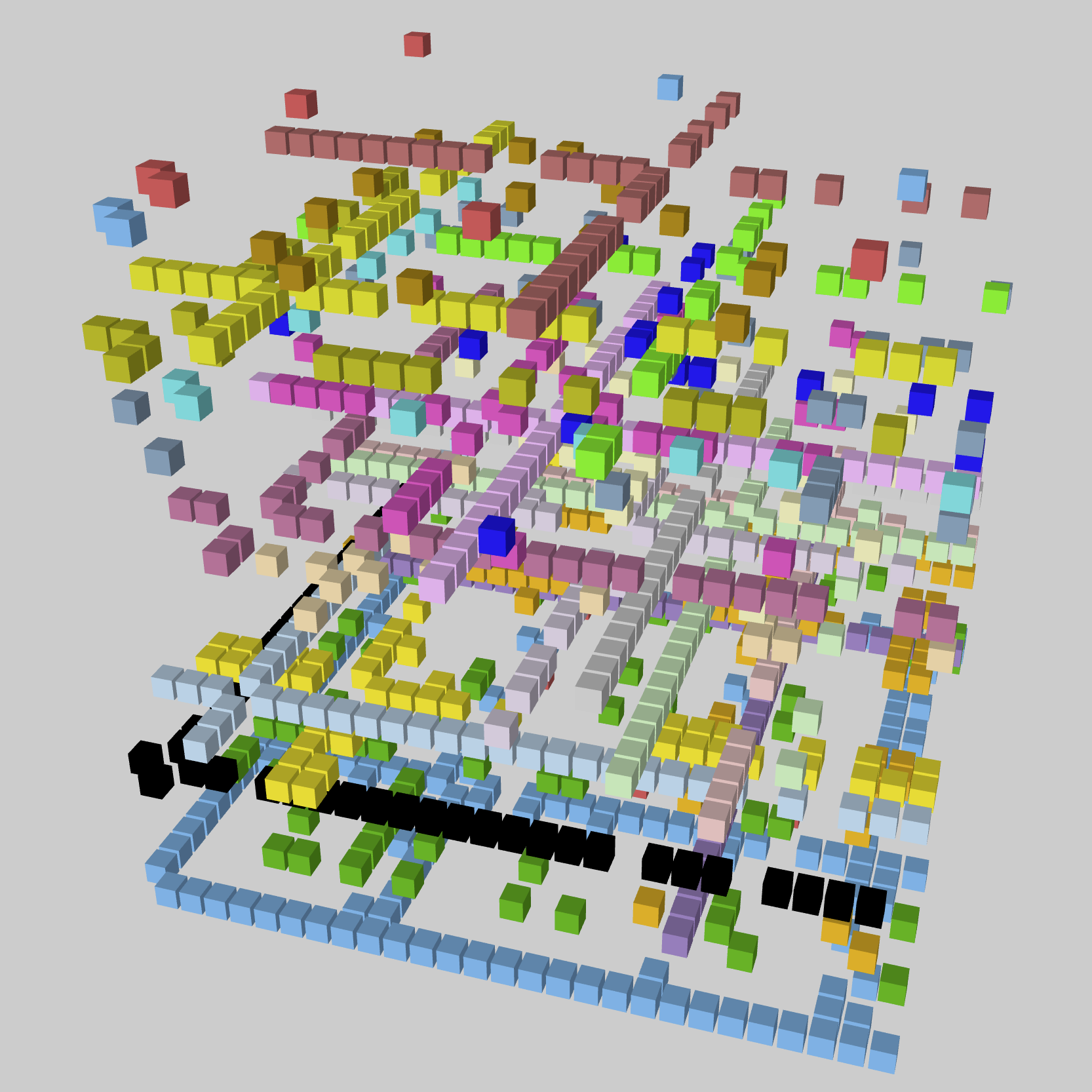} 
\includegraphics[width=62mm]{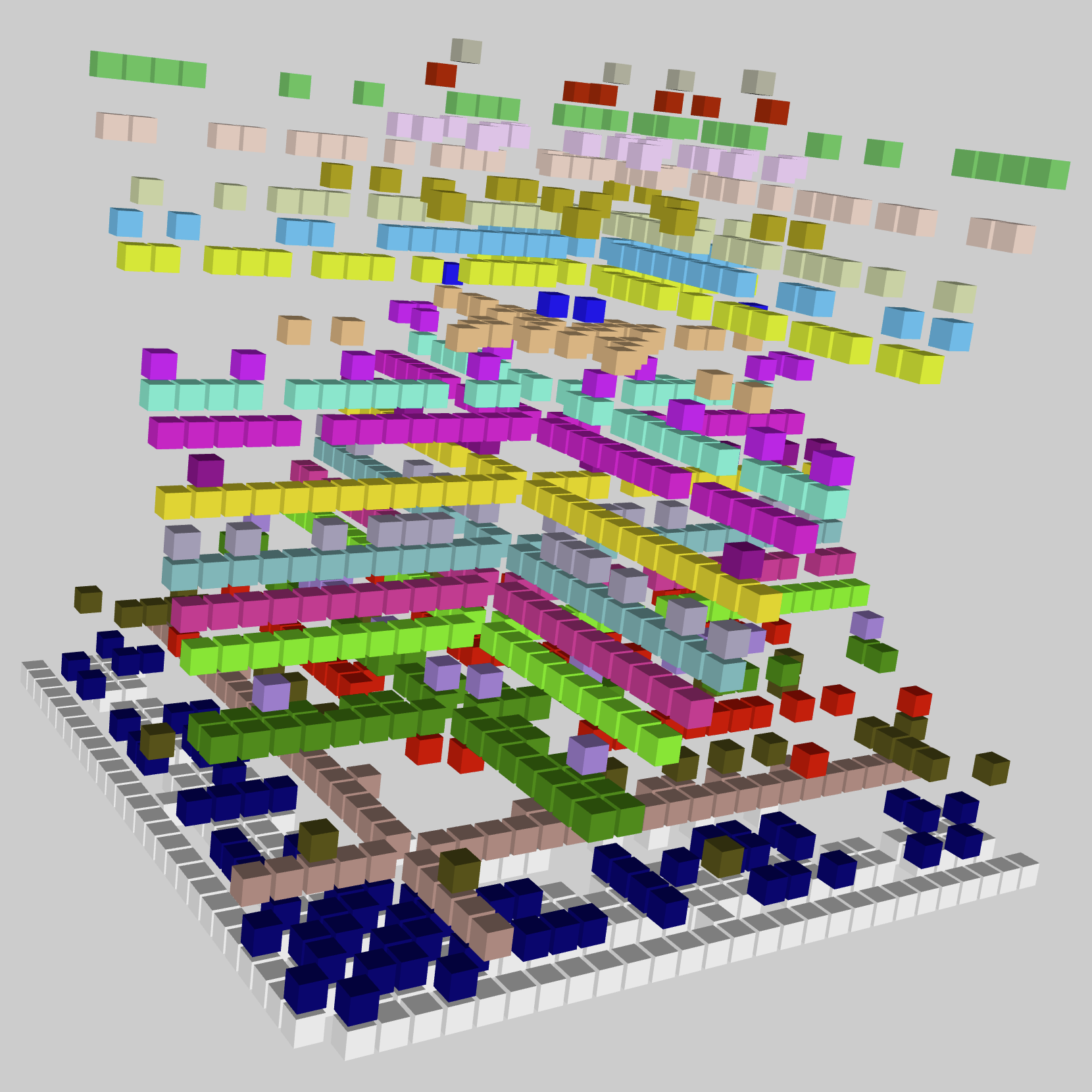}
\caption{3D layout of EUair diversity core at level 13; initial (left) and reordered by crosses (right).\label{EUcore}}
\end{figure}

For visual inspection we created a 3D layout of the core in X3D format -- see the left side picture in Figure~\ref{EUcore}. 
The layout contains many "crosses" -- lines from a company's base airport. As the first attempt, we reordered companies by the number of lines and afterward the airports by the main airport for each company.

\begin{figure}
\includegraphics[width=62mm]{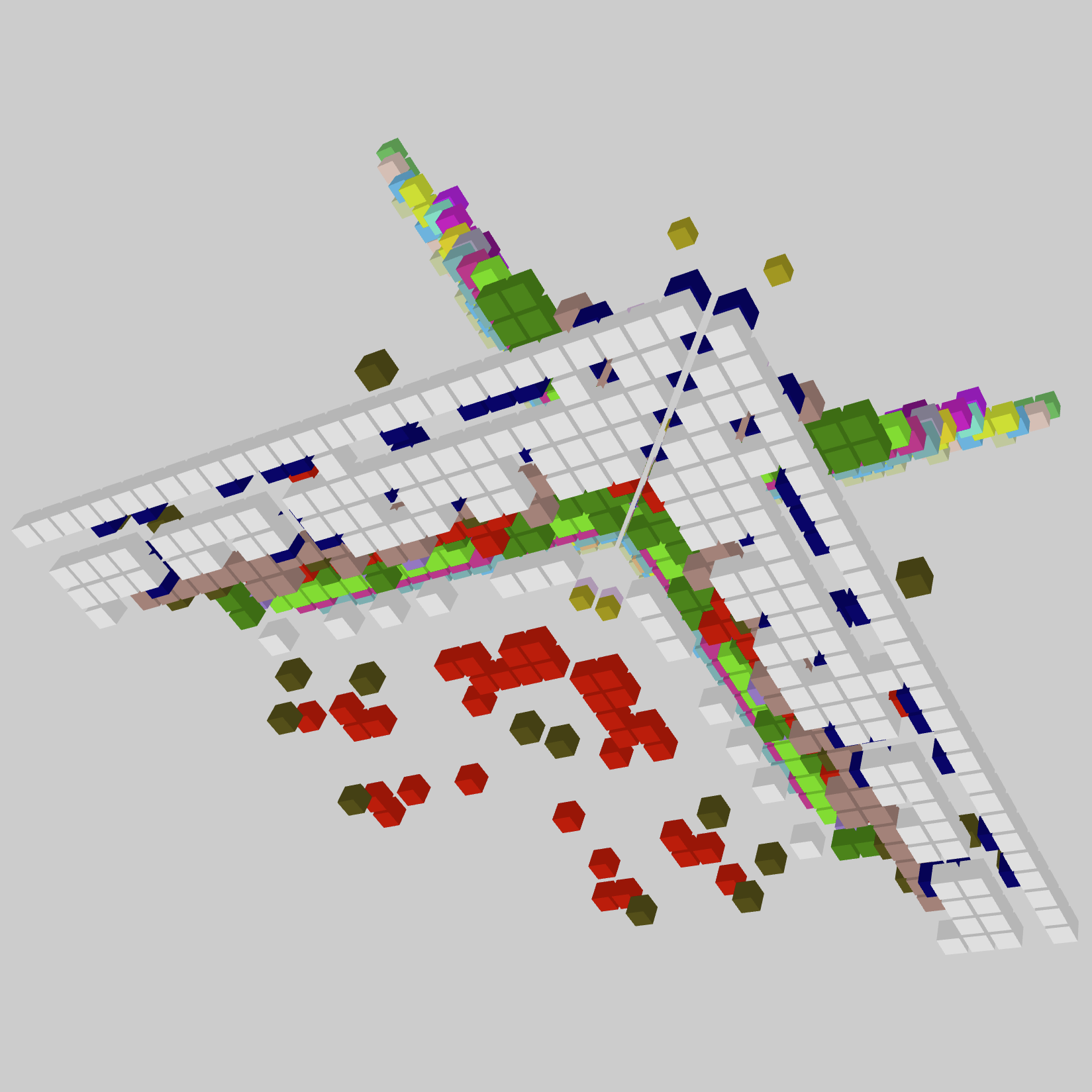}
\includegraphics[width=62mm]{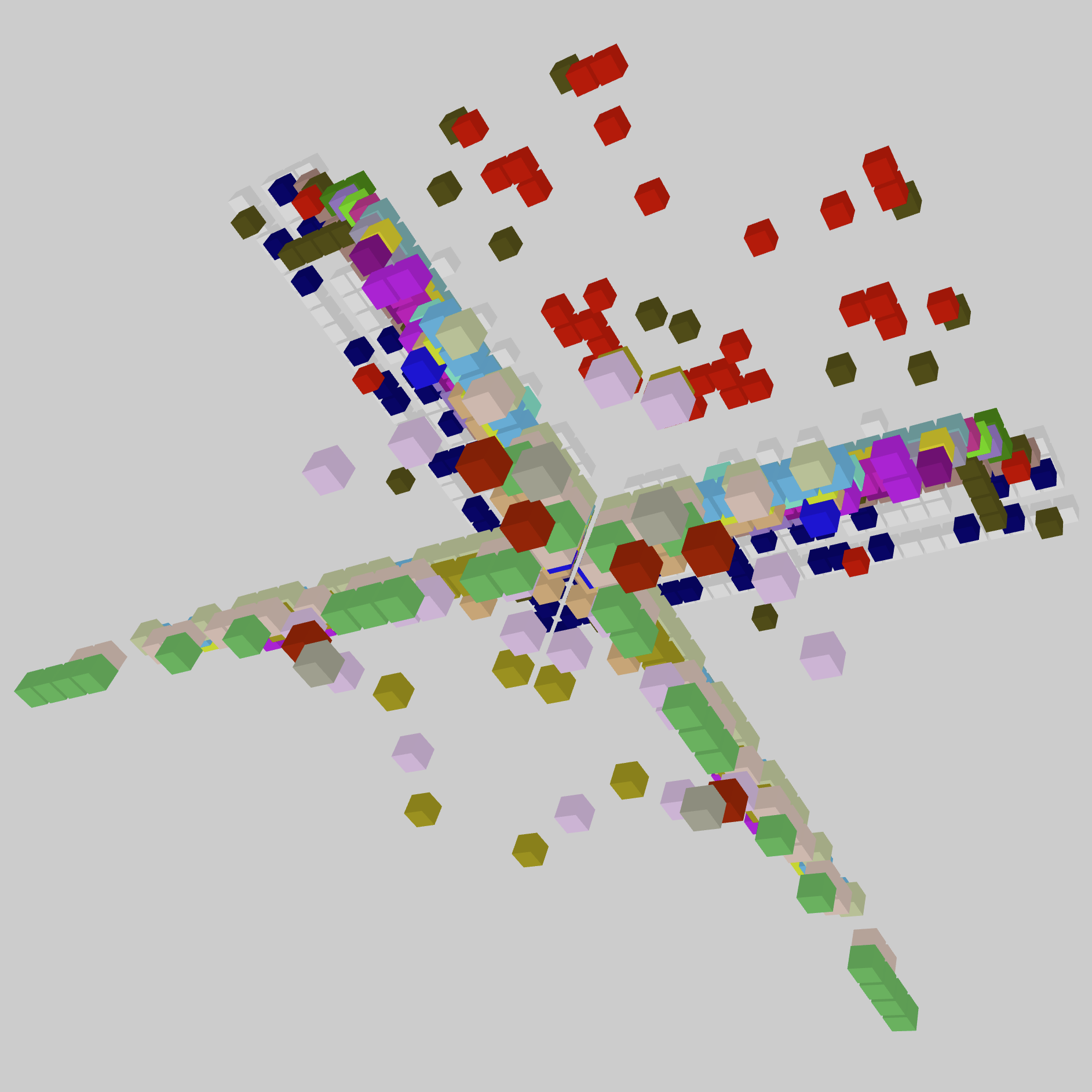}
\caption{3D layout of EUair diversity core at level 13; two views exposing companies based on multiple airports.\label{EUcord}}
\end{figure}

The obtained layout confirmed our observation -- most companies are providing lines from their base airport. It can be further improved by reordering (making closer) companies having the same base airport or airports served by the same company -- see the right side figure.
The bottom (left) and top (right) side views in Figure~\ref{EUcord} show that most of the links lie on crosses, but also that there are some companies with "dispersed" lines linking different airports.

Inspecting the layout in detail we can make the following observations about the core:

\begin{itemize}
\item Lufthansa and Air Berlin are serving lines from German airports Frankfurt, Berlin, Munich, Dusseldorf, Hamburg, and Milan in Italy
\item Swiss IAL and Netjets are serving lines from Swiss airports Zurich and Geneva
\item SAS and Norwegian AS are serving lines from Copenhagen and Stockholm
\item Spanish airports Madrid, Barcelona, and Malaga are served by companies Iberia, Air Nostrum, Vueling, and Ryanair
\item companies serving lines from their base airport: (British A, Heathrow), (LOT, Warsaw), (Austrian A, Niki; Vienna), (KLM, Transavia; Amsterdam), (Air France, CDG), (Czech A, Prague), (Alitalia, Fiumicino), (Brussels A, European AT; Brussels), (Malev, Budapest), (Aegean, Olympic Air; Athens); the only “irregular” link is by Iberia between Barcelona and Budapest
\item Companies with dispersed services inside the core are Easyjet, Netjets, Wizz Air, European AT, and TNT Airways
\begin{itemize}
\item Easyjet is serving lines from Milan and also from CDG, Amsterdam, Fiumicino, Madrid
\item Netjets is serving lines also from Barcelona, Venice, Vienna, and Nice
\item Wizz Air is serving lines from Budapest and Fiumicino
\item European AT has also some lines from Heathrow, Milan, Venice, Barcelona, and Madrid
\item TNT Airways is linking airports CDG and Athens, and Sofia and Henri Coanda
\end{itemize}
\end{itemize}

\subsection{Summer Olympic medals till 2016}

The data set \emph{120 years of Olympic history -- athletes and results} with basic bio data on athletes and medal results from Athens 1896 to Rio 2016 is available at Kaggle \cite{olymp}. The original data describe 134732 participants of the Summer and Winter Olympics. We transformed it into the multiway network \texttt{Olympics16S} about the medalists of the Summer Olympics till 2016 (available at GitHub/Bavla \cite{MWnets}). It contains data about 34088 medals. Omitting the athlete Name info and simplifying the network we obtain a multiway network with 10429 links on five ways
(Games (29), NOC (147), Sport (52), Sex (2), Medal (3)), the weight $w$ counting the medals, and additional weights Age, Height, and Weight.  

We decided to search for a core with a large number of medals on the ways Games, NOC, and Sport.

After examing the distributions of values of the node property function $p_S$ on all three ways we decided to use the core conditions list $P$ that contains three triples $(p=pW\!sum, t=100, args=(Games, NOC, Sport) )$, $(p=pW\!sum, t=100, args=(NOC, Games, Sport) )$,  and $(p=pW\!sum, t=100, args=(Sport, Games, NOC) )$. The obtained core was too large. We increased the thresholds to (500, 300, 350). The corresponding $p_S$-core  $\C = (C_{\mathrm{Games}}, C_{\mathrm{NOC}}, C_{\mathrm{Sport}})$ is of order (577, 308, 390) -- inside the core, each Olympics accounts for at least 577 medals, each country (NOC) for at least 308 medals, and each sport discipline for at least 390 medals.\medskip

\noindent
$C_{\mathrm{Games}}$ (25):
1988 Seoul (1405), 1984 Los Angeles (1354), 2008 Beijing (1352), 2012 London (1338), 2000 Sydney (1324), 
2016 Rio de Janeiro (1320), 2004 Athens (1305), 1980 Moscow (1270), 1996 Atlanta (1251), 1976 Montreal (1220), 
1992 Barcelona (1112), 1972 Munich (1080), 1920 Antwerp (1054), 1964 Tokyo (952), 1968 Mexico City (949), 
1912 Stockholm (853), 1960 Rome (826), 1952 Helsinki (797), 1956 Melbourne $+$ Stockholm (795), 1936 Berlin (754), 
1948 London (731), 1924 Paris (681), 1908 London (656), 1928 Amsterdam (610), 1932 Los Angeles (577).

\noindent
$C_{\mathrm{NOC}}$ (29):
 USA (4153), URS (2027), GBR (1594), GER (1573), ITA (1325), FRA (1184), AUS (1165), HUN (1052), SWE (1019),
 NED (854), GDR (843), RUS (776), CHN (662), JPN (661), ROU (632), CAN (613), NOR (546), POL (520), DEN (519), 
 FRG (500), FIN (446), BRA (446), ESP (416), YUG (379), SUI (352), KOR (347), BEL (337), BUL (317), TCH (308).

\noindent
$C_{\mathrm{Sport}}$ (21):  
Athletics (2911), Swimming (2815), Rowing (2535), Gymnastics (2043), Fencing (1584), Football (1216), Cycling (1129),
Hockey (1126), Canoeing (1010), Shooting (995), Sailing (995), Wrestling (941), Handball (924), Basketball (914),
Equestrianism (883), Water Polo (876), Volleyball (839), Boxing (596), Judo (425), Weightlifting (419), Diving (390).

\begin{figure}
\includegraphics[height=53mm]{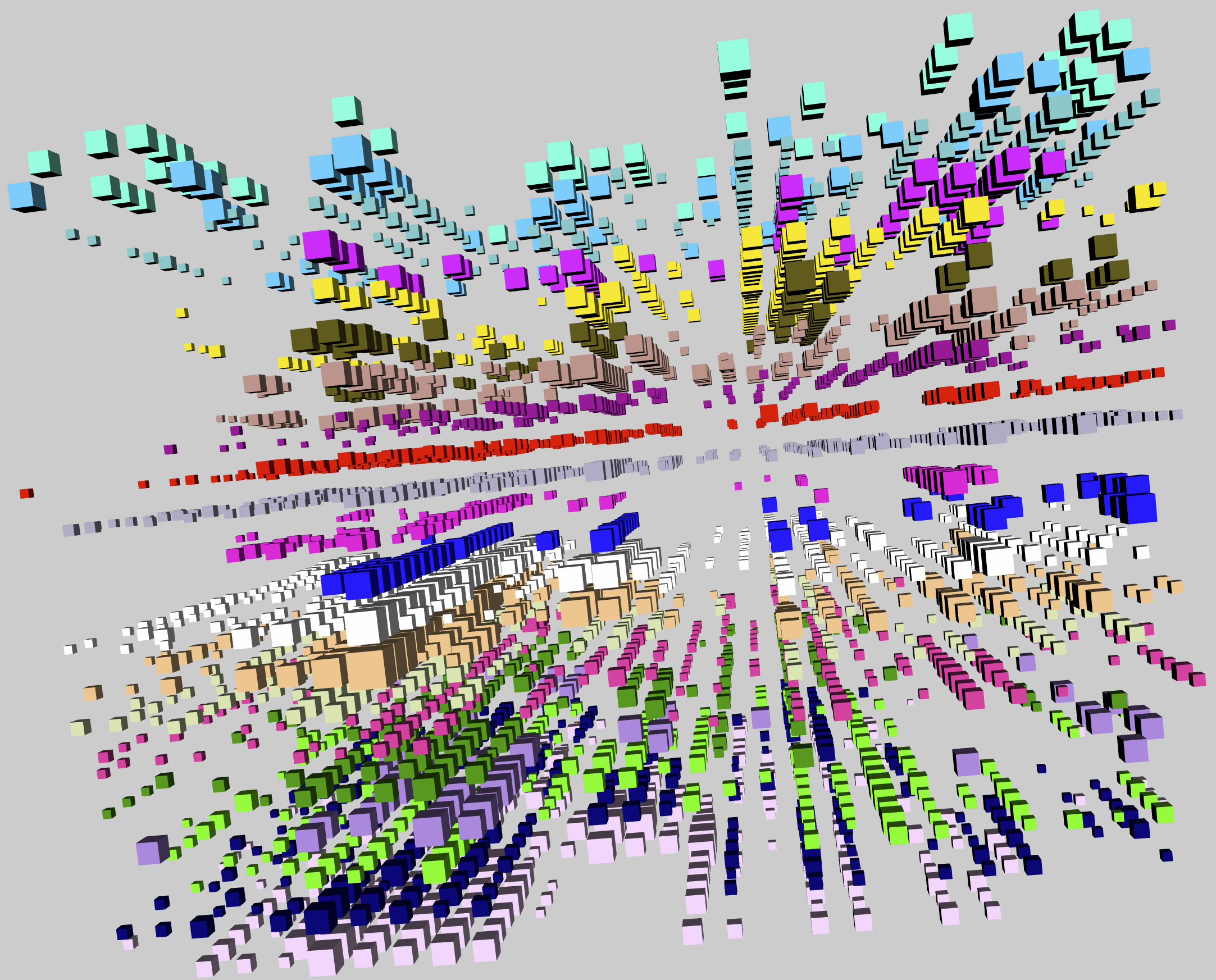} \  \includegraphics[height=53mm]{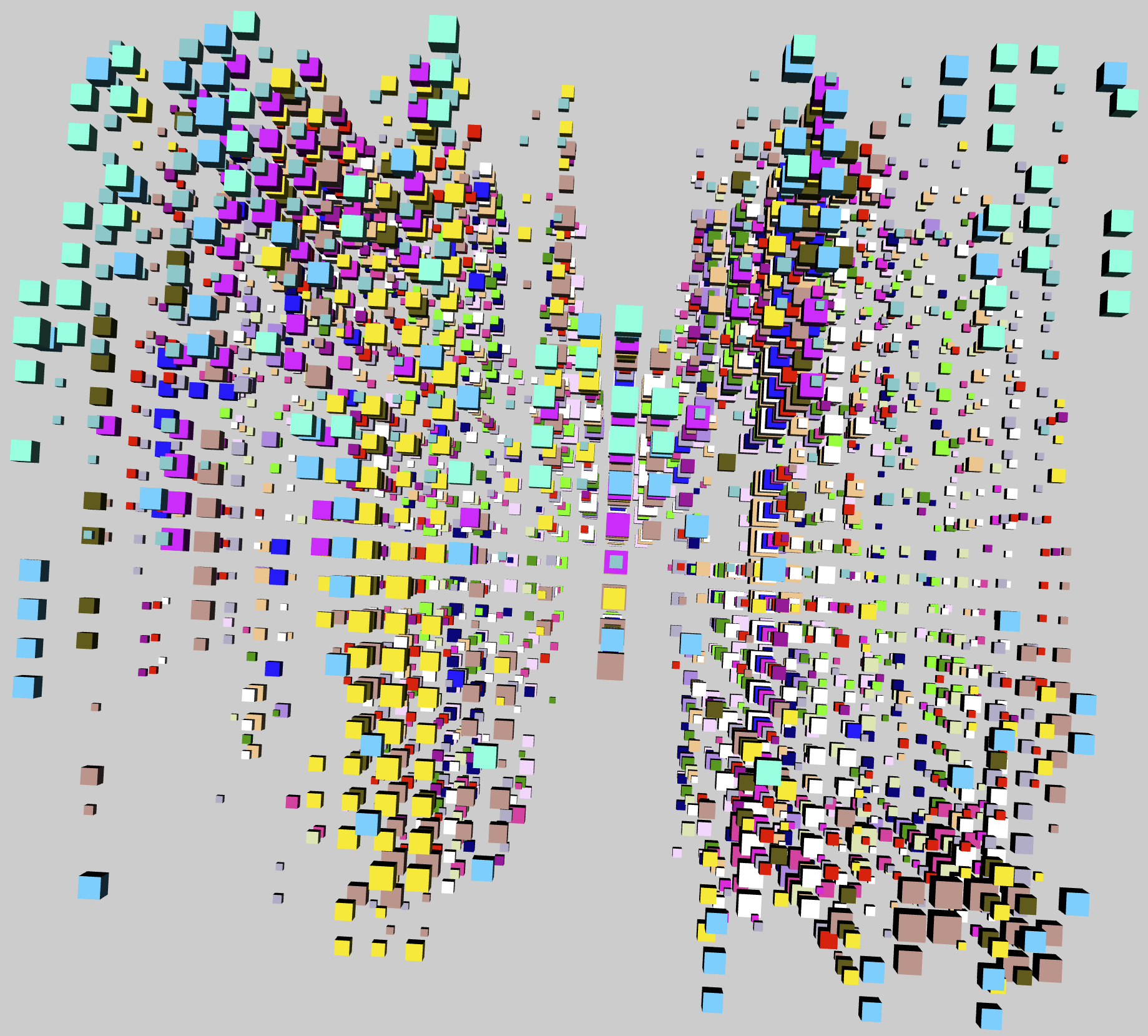}
\caption{Screenshot of a 3D layout of Olympics weighted degree core at levels (577, 308,390).\label{OLcore}}
\end{figure}

It turns out that the core contains $25566/34088 = 75 \%$ of all medals, $3607/10429 = 34.6 \%$ of all links, and occupies $6.9 \%$ of the total space. Note also that the set $\C_{\mathrm{NOC}}$ contains the Soviet Union and Russia; and Germany, West Germany, and East Germany.

We produced a 3D layout for the core subnetwork of \texttt{Olympics1}. See Figure~\ref{OLcore}. The links belonging to the same sport discipline are of the same color. The Olympics are ordered chronologically. The order of NOCs (countries) and sport disciplines was determined using clustering.

Some observations:
\begin{itemize}
\item "Towers" in the same color (sport discipline) slices represent countries that were dominant in these disciplines through games. For example, the ochre, white and blue towers in the top mid of the bottom left quadrant in the left picture of Figure~\ref{OLcore} represent USA swimming, athletics, and basketball. A detailed inspection reveals
missing cubes in 1980 (boycott of Moscow Olympics). The pink slice at the bottom represents rowing and has towers for NOCs GBR, AUS, USA, GER, NED, FRA, ITA, DEN, and NOR. 
\item The holes at the bottom and the top in the middle of the right picture of Figure~\ref{OLcore} belong to NOCs  FRG, URS, and GDR. The hole on the right in the middle belongs to Germany (GER). 
\item The three yellow (fencing) towers on the right picture of Figure~\ref{OLcore} belong to HUN, ITA, and FRA.
\item The holes in another view correspond to the introduction of new sports disciplines:  canoeing, basketball (1936), judo, and volleyball (1964). It seems that handball was first played in Berlin in 1936 and was reintroduced in Munich in 1972.
\end{itemize}

\section{Conclusions}

In this paper, we extended the notion of the generalized core to multiway networks and we proposed a new node property function -- diversity of a node.

We also developed an algorithm for determining the generalized core in a given multiway network for selected node properties and corresponding threshold levels. The algorithm is very general -- it should work for any selection of monotonic node property functions. For very large networks and special node property functions much more efficient algorithms can be developed following the same basic idea. For inspection of the obtained results, we developed the function that for a given multiway network creates its 3D layout and exports it in X3D format.

The multiway network core algorithm, some node property functions, and the creation of 3D layouts are included in our R package MWnets for multiway network analysis. It is still under development. Its current version is available at Github/Bavla \cite{MWnets}. In the future, we intend to program it also in Python or/and Julia.

One of the important aspects of developing new data analytic methods is having a collection of interesting data sets as a source of problems/ideas, test cases, and (understandable) illustrations. Such a collection of multiway networks is also available at Github/Bavla.

Our application of the multiway network core algorithm to selected data sets (European airports and Olympic medals), allowed us to identify the most cohesive subnetworks in both networks and by their inspection reveal interesting observations about their structure. An interesting question for further research is the normalization of multiway networks. We also intend to investigate what happens in the multiway networks in which some ways were obtained by discretization of numerical variables (conversion of numerical scale into nominal scale).

\section*{Acknowledgments}

We would like to thank Andreas Plesch for extending the library \texttt{X3DOM} with support for \textbf{\texttt{Anchor}}'s attribute \textbf{\texttt{description}}.

\ifblind \else
The computational work reported in this presentation was performed using R library MWnets. The code and data are available at Github/Bavla \cite{MWnets}. 

This work is supported in part by the Slovenian Research Agency
 (research program P1-0294 and research projects J5-2557, J1-2481 and J5-4596),
 and prepared within the framework of the COST action CA21163 (HiTEc).
\fi



\end{document}